\documentclass{article}
\usepackage{graphicx,amsmath,algorithmic,algorithm,amssymb,subcaption,xspace,amsthm,multicol}
\usepackage{fullpage}
\pdfoutput=1

\newcommand{\eg}{e.g.\xspace}

\newcommand{\Q}{\mathcal{Q}}
\newcommand{\K}{\mathcal{K}}
\newcommand{\saemQ}{\widehat{\mathcal{Q}}}
\newcommand{\saemS}{\mathbb{S}}
\newcommand{\T}{T}
\newcommand{\myd}{\textrm{d}}      
\newcommand{\nx}{n_z} 
\newcommand{\nuu}{n_u} 
\newcommand{\ny}{n_y} 
\newcommand{\argmax}[2]{\underset{#1}{\mathrm{arg}\,\mathrm{max}}\,\,#2}
\newcommand{\myDirac}[2]{\delta_{#1}(#2)}
\newcommand{\Np}{N}

\newcommand{\rev}[1]{#1}
\DeclareMathOperator{\Tr}{Tr}

\newtheorem{lemma}{Lemma}
\newcommand{\ls}{z}
\newcommand{\nls}{s}

\newcommand{\Expb}[2]{\operatorname{\mathbb{E}}_{#1}\left[#2\right]}
\newcommand{\Ind}[1]{\operatorname{\boldsymbol{1}}\left(#1\right)}
\newcommand{\Cov}[1]{\operatorname{Cov}\left[#1\right]}

\newcommand{\Prob}[1]{\mathbb{P}\left(#1\right)}

\title{
Identification of jump Markov linear models using particle filters
}

\author{Andreas Svensson, Thomas B. Sch\"{o}n and Fredrik Lindsten\footnote{
This work was supported by the project \emph{Probabilistic modelling of dynamical systems} (Contract number: 621-2013-5524) funded by the Swedish Research Council (VR) and  the project \emph{Learning of complex dynamical systems} (Contract number: 637-2014-466) funded by the Swedish Research Council (VR). Andreas Svensson and Thomas B. Sch\"{o}n are with the Department of Information Technology, Uppsala University, Sweden {\tt\small \{andreas.svensson, thomas.schon\}@it.uu.se}, and Fredrik Lindsten is with the Department of Engineering, University of Cambridge, UK {\tt\small fredrik.lindsten@eng.cam.ac.uk}
}
}

\begin{document}

\maketitle
\thispagestyle{empty}
\pagestyle{empty}

\begin{abstract}
Jump Markov linear models consists of a finite number of linear state space models and a discrete
variable encoding the jumps (or switches) between the different linear models. Identifying jump
Markov linear models makes for a challenging problem lacking an analytical solution. We derive a new
expectation maximization (EM) type algorithm that produce maximum likelihood estimates of the
model parameters. Our development hinges upon recent progress in combining particle filters with
Markov chain Monte Carlo methods in solving the nonlinear state smoothing problem inherent in the
EM formulation. Key to our development is that we exploit a conditionally linear Gaussian
substructure in the model, allowing for an efficient algorithm.

\end{abstract}

\section{Introduction}

Consider the following \emph{jump Markov linear model} on state space form
\begin{subequations}
  \label{eq:jmlgss}
  \begin{align}
    \nls_{t+1}\mid \nls_t & \sim p(\nls_{t+1}|\nls_t), \\
    \ls_{t+1} &= A_{\nls_{t+1}}\ls_t + B_{\nls_{t+1}}u_t + w_t, \\
    y_t &= C_{\nls_{t}}\ls_t + D_{\nls_{t}}u_t + v_t,
  \end{align}
\end{subequations}
where $\sim$ means distributed according to and the (discrete) variable $\nls_t$ takes values in $\{
1, \dots, K \}$ (which can be thought of as different \emph{modes} which the model is \emph{jumping} between) and the
(continuous) variable $\ls_t$ lives in $\mathbb{R}^{\nx}$. Hence, the state variable consists of
$x_t \triangleq \left( \ls_t, ~ \nls_t \right)$. Furthermore, $v_t \in
\mathbb{R}^{\ny}$ and $w_t \in \mathbb{R}^{\nx}$ are zero mean white Gaussian noise and $\mathbb{E}
w_tw_t^T = Q_{\nls_{t+1}}$, $\mathbb{E} v_tv_t^T = R_{\nls_t}$ and $\mathbb{E} w_tv_t^T \equiv 0$.
The output (or measurement) is $y_t \in \mathbb{R}^{\ny}$, the input is $u_t \in
\mathbb{R}^{\nuu}$. As $K$ is finite, $p(\nls_{t+1}|\nls_t)$ can be defined via a matrix $\Pi \in
\mathbb{R}^{K\times K}$ with entries $\pi_{mn} \triangleq p(\nls_{t+1}=n|\nls_t = m)$. 

We are interested in off-line identification of jump Markov linear models on the
form~\eqref{eq:jmlgss} for the case of an \emph{unknown jump sequence}, but the number of
modes $K$ is known. More specifically, we will formulate and solve the Maximum Likelihood (ML)
problem to compute an estimate of the static parameters $\theta$ of a jump Markov linear model based
on a batch of measurements $y_{1:T} \triangleq \{y_1, \dots, y_{\T}\}$ and (if available) inputs
$u_{1:T}$ by solving,
\begin{align}
  \label{eq:ML}
  \widehat{\theta}_{\text{ML}} = \argmax{\theta \in \Theta} p_{\theta}(y_{1:T}).
\end{align}
Here $\theta \triangleq \{ \{ A_n , B_n, C_n, D_n , Q_n, R_n \}_{n=1}^K, \Pi \}$, i.e., all unknown
static parameters in model~\eqref{eq:jmlgss}. Here, and throughout the paper, the dependence on the inputs $u_{1:T}$ is implicit.

Solving~\eqref{eq:ML} is challenging and there are no closed form solutions available. Our approach
is to derive an expectation maximization (EM)~\cite{dempster1977maximum} type of solution, where the
strategy is to separate the original problem into two closely linked problems. The first problem is
a challenging, but manageable nonlinear state smoothing problem and the second problem is a
tractable optimization problem. The nonlinear smoothing problem we can solve using a combination of
sequential Monte Carlo (SMC) methods (particle filters and particle smoothers)
\cite{doucet2009tutorial} and Markov chain Monte Carlo (MCMC) methods \cite{Robert2004mcs}. More
specifically we will make use of particle MCMC (PMCMC), which is a systematic way of exploring the
strengths of both approaches by using SMC to construct the necessary high-dimensional Markov
kernels needed in MCMC~\cite{andrieu2010particle,lindsten2014particle}.

Our main contribution is a new maximum likelihood estimator that can be used to identify jump Markov
linear models on the form~\eqref{eq:jmlgss}. The estimator exploits the conditionally linear
Gaussian substructure that is inherent in~\eqref{eq:jmlgss} via Rao-Blackwellization. More
specifically we derive a Rao-Blackwellized version of the particle stochastic approximation expectation maximization (PSAEM)
algorithm recently introduced in~\cite{lindsten2013efficient}.

Jump Markov linear models, or switching linear models, is a fairly well studied class of hybrid
systems. For recent overviews of existing system identification
methods for jump Markov linear models, see \cite{garulli2012survey,
  paoletti2007identification}. Existing approaches considering the problem under study here include
two stage methods, where the data is first segmented (using \eg change detection type of methods)
and the individual models are then identified for each segment, see \eg
\cite{pekpe2004identification, borges2005switching}.
There has also been approximate EM algorithms
proposed for identification of hybrid systems~\cite{blackmore2007model, gil2009beyond} and the very recent~\cite{ashley2014sequential}
(differing from our method in that we use stochastic approximation EM and Rao-Blackwellization). There are
also relevant relationships to the PMCMC solutions introduced in~\cite{whiteley2010efficient} and
the SMC-based on-line EM solution derived in~\cite{yildirim2013online}.

There are also many approaches considering the more general problem with an unknown number of modes $K$
and an unknown state dimension $n_\ls$, see \eg \cite{fox2011bayesian} and
\cite{bemporad2001identification}, making use of Bayesian nonparametric models and mixed integer
programming, respectively.

\section{Expectation maximization algorithms}
The EM algorithm \cite{dempster1977maximum} provides an iterative method for
computing maximum likelihood estimates of the unknown parameters $\theta$ in a probabilistic model
involving latent variables. In the jump Markov linear model~\eqref{eq:jmlgss} we observe $y_{1:\T}$,
whereas the state $x_{1:\T}$ is latent.

The EM algorithm maximizes the likelihood by iteratively maximizing the \emph{intermediate quantity}
\begin{align}
  \Q(\theta, \theta^{\prime}) &\triangleq \int \log p_{\theta}(x_{1:\T}, y_{1:\T}) p_{\theta^{\prime}}(x_{1:\T}\mid y_{1:\T})\myd x_{1:\T}. \label{eq:EM:Qdef}
\end{align}
More specifically, the procedure is initialized in $\theta_0\in\Theta$ and then iterates between
computing an expected (E) value and solving a maximization (M) problem,
\begin{align*}
  \text{(E) }\, &\text{Compute } \Q(\theta,\theta_{k-1}).\\
  \text{(M) }\, &\text{Compute } \theta_{k} = \argmax{\theta\in\Theta}{\Q(\theta, \theta_{k-1})}.
\end{align*}
Intuitively, this can be thought of as `selecting the new parameters as the ones that make the given measurements \emph{and} the current state estimate as likely as possible'.

The use of EM type algorithms to identify dynamical systems is by now fairly well explored for both
linear and nonlinear models. For linear models, there are explicit expressions for all involved
quantities, see e.g. \cite{gibson2005robust,shumway1982approach}.  For nonlinear models the
intermediate quantity $\Q(\theta,\theta^{\prime})$ is intractable and we are forced to approximate
solutions; see e.g. \cite{lindsten2013efficient, SchonWN:2011, OlssonDCM:2008, CappeMR:2005}. This
is the case also for the model~\eqref{eq:jmlgss} under study in this work. Indeed, the maximization step can be solved in closed form for the model \eqref{eq:jmlgss},
but~\eqref{eq:EM:Qdef} is still intractable in our case.

It is by now fairly well established that we can make use of sequential Monte Carlo (SMC) \cite{doucet2009tutorial} or particle Markov chain Monte Carlo (PMCMC) \cite{andrieu2010particle} methods to approximate the joint smoothing distribution
for a general nonlinear model arbitrarily well according to
\begin{align}
  \label{eq:PSA}
  \widehat{p}(x_{1:\T}\mid y_{1:\T}) = \sum_{i=1}^{\Np}w_\T^{i}\myDirac{x_{1:\T}^{i}}{x_{1:\T}},
\end{align}
where $x_{1:T}^i$ are random samples with corresponding importance weights $w_T^i$, $\delta_x$ is a point-mass distribution at $x$ and we refer to $\{x_{1:\T}^{i}, w_\T^{i}\}_{i=1}^{\Np}$ as a \emph{weighted particle system}.
The particle smoothing approximation~\eqref{eq:PSA} can be used to approximate the integral in~\eqref{eq:EM:Qdef}.
Using this approach within EM, we obtain the particle smoothing EM (PSEM) method~\cite{OlssonDCM:2008,SchonWN:2011}. PSEM can be viewed as an SMC-analogue of the well known Monte Carlo EM (MCEM) algorithm \cite{wei1990monte}.

However, it has been recognized that MCEM, and analogously PSEM, makes inefficient use of the generated samples \cite{delyon1999convergence}. This is particularly true when the simulation step is computationally expensive,
which is the case when using SMC or PMCMC. To address this shortcoming, \cite{delyon1999convergence}
proposed to use a \emph{stochastic approximation}  (SA) \cite{robbins1951stochastic} of the intermediate quantity instead of a vanilla Monte Carlo approximation, resulting in the stochastic approximation EM (SAEM) algorithm. The SAEM algorithm replaces the intermediate quantity $\Q$ in EM with
\begin{align}
  \label{eq:Qsaem}
  \saemQ_k(\theta) = (1-\gamma_k)\saemQ_{k-1}(\theta) + \gamma_k \log p_{\theta}(y_{1:T}, x_{1:T}[k]),
\end{align}
with $\{\gamma_k\}_{k=1}^\infty$ being a sequence of step sizes which fulfils $\sum_{k=1}^\infty \gamma_k = \infty$ and $\sum_{k=1}^\infty \gamma_k^2 < \infty$. In the above,
$x_{1:T}[k]$ is a sample state trajectory, simulated from the joint smoothing distribution $p_{\theta_k}(x_{1:T}\mid y_{1:T})$.
It is shown by \cite{delyon1999convergence} that the SAEM algorithm---which iteratively updates the intermediate
quantity according to \eqref{eq:Qsaem} and computes the next parameter iterate by maximizing this stochastic approximation---enjoys good convergence properties. Indeed, despite the fact that the method requires
only a single sample $x_{1:T}[k]$ at each iteration, the sequence $\{\theta_k\}_{k\geq1}$ will converge
to a maximizer of $p_\theta(y_{1:T})$ under reasonably weak assumptions.

However, in our setting it is not possible to simulate from the joint smoothing distribution
$p_{\theta_k}(x_{1:T}\mid y_{1:T})$.  We will therefore make use of the particle
SAEM (PSAEM) method \cite{lindsten2013efficient}, which combines recent PMCMC methodology with
SAEM. Specifically, we will exploit the structure of~\eqref{eq:jmlgss} to develop a
Rao-Blackwellized PSAEM algorithm.

We will start our development in
the subsequent section by considering the smoothing problem for~\eqref{eq:jmlgss}. We derive
a PMCMC-based Rao-Blackwellized smoother for this model class.  The proposed smoother can,
principally, be used to compute \eqref{eq:EM:Qdef} within PSEM.  However, a more efficient approach is to use the proposed
smoother to derive a Rao-Blackwellized PSAEM algorithm, see Section~\ref{sec:identification}.

\section{Smoothing using Monte Carlo methods}
For smoothing, that is, finding $p_{\theta}(x_{1:t}|y_{1:t}) = p_{\theta}(\nls_{1:T}, \ls_{1:T}|y_{1:T})$, various Monte Carlo methods can be applied.
We will use an MCMC based approach, as it fits very well in the SAEM framework (see \eg \cite{andrieu2005stability, kuhn2004coupling}), which together shapes the PSAEM algorithm. The aim of this section is therefore to derive an MCMC-based smoother for jump Markov linear models.

To gain efficiency, the
jump sequence $\nls_{1:\T}$ and the linear states $\ls_{1:\T}$ are separated using
conditional probabilities as
\begin{align}
p_{\theta}(\nls_{1:T}, \ls_{1:T}|y_{1:T}) = p_{\theta}(\ls_{1:T}|\nls_{1:T}, y_{1:T})p_{\theta}(\nls_{1:T}|y_{1:T}). \label{eq:rb}
\end{align}
This allows us to infer the
conditionally linear states $\ls_{1:\T}$ using closed form expressions. Hence, it is only the jump
sequence $\nls_{1:\T}$ that has to be computed using approximate inference. This technique is
referred to as Rao-Blackwellization~\cite{casella1996rao}.

\subsection{Inferring the linear states: $p(\ls_{1:T} | \nls_{1:T}, y_{1:T})$}\label{sec:lgss}

State inference in linear Gaussian state space models can be performed exactly in closed form. More
specifically, the Kalman filter provides the expressions for the filtering PDF
$p_\theta(\ls_t|\nls_{1:t},y_{1:t}) = \mathcal{N}(\ls_t|\widehat{\ls}_{f;t},P_{f;t})$ and the one
step predictor PDF $p_\theta(\ls_{t+1}|\nls_{1:t+1},y_{1:t}) =
\mathcal{N}(\ls_t|\widehat{\ls}_{p;t+1},P_{p;t+1})$. The marginal smoothing PDF
$p_\theta(\ls_{t}|\nls_{1:T},y_{1:T}) = \mathcal{N}(\ls_t|\widehat{\ls}_{s;t},P_{s;t})$ is provided
by the Rauch-Tung-Striebel (RTS) smoother~\cite{rauch1965maximum}. See, e.g., \cite{kailath2000linear} for the relevant results. Here, we use
$\mathcal{N}(x\mid \mu, \Sigma)$ to denote the PDF for the (multivariate) normal distribution with
mean $\mu$ and covariance matrix $\Sigma$.

\subsection{Inferring the jump sequence: $p(\nls_{1:T} | y_{1:T})$}\label{sec:nl}

To find $p(\nls_{1:T} | y_{1:T})$, an MCMC approach is used. First, the concept of using Markov kernels for smoothing is introduced, and then the construction of the kernel itself follows.

MCMC makes use of ergodic theory for statistical inference.
Let $\K_\theta$ be a Markov kernel (to be defined below) on the $T$-fold product space
$\{1,...,K\}^T$. Note that the jump sequence $\nls_{1:T}$ lives in this space.
Furthermore, assume that $\K_\theta$ is \emph{ergodic} with unique stationary
distribution $p_\theta(\nls_{1:T} | y_{1:T})$.
This implies that by simulating a Markov chain with transition kernel $\K_\theta$, the marginal distribution
of the chain will approach $p_\theta(\nls_{1:T} | y_{1:T})$ in the limit.

Specifically, let $\nls_{1:T}[0]$ be an arbitrary initial state with $p_\theta( \nls_{1:T}[0] | y_{1:T}) > 0$ and let $\nls_{1:T}[k] \sim \K_\theta(\cdot | s_{1:T}[k-1]))$ for $k \geq 1$, then by the ergodic theorem \cite{Robert2004mcs}:
\begin{align}
\frac{1}{n}\sum_{k=1}^n h(s_{1:T}[k]) \rightarrow \Expb{\theta}{ h(s_{1:T}) | y_{1:T} }, \label{eq:Kmcmc}
\end{align}
as $n \rightarrow \infty$ for any function $h: \{1,...,K\}^T \mapsto \mathbb{R}$. This allows a smoother to be constructed as
in Algorithm~\ref{alg:rbpmcmc}.

\begin{algorithm}
\caption{MCMC smoother}
\label{alg:rbpmcmc}
\begin{algorithmic}[1]
\STATE Initialize $\nls_{1:T}[0]$ arbitrarily
\FOR{$k \geq 1$}
\STATE Generate $\nls_{1:T}[k] \sim \K_\theta(\cdot|\nls_{1:T}[k-1])$
\ENDFOR
\end{algorithmic}
\end{algorithm}

We will use the \emph{conditional particle filter with ancestor sampling} (CPF-AS) \cite{lindsten2014particle} to construct the Markov kernel $\K_\theta$. 
The CPF-AS is similar to a standard particle filter, but with the important difference that one particle trajectory (jump sequence), $\nls'_{1:T}$, is specified \emph{a priori}.

The algorithm statement for the CPF-AS can be found in, e.g., \cite{lindsten2014particle}. Similar to an auxiliary particle filter \cite{doucet2009tutorial}, the propagation of $p_\theta(\nls_{1:t-1}|y_{1:t-1})$ (approximated by $\{\nls_{1:t-1}^{i}, w_{t-1}^{i}\}_{i=1}^{\Np}$) to time $t$ is done using the \emph{ancestor indices} $\{ a_t^i\}_{i=1}^N$. To generate $\nls_t^i$, the ancestor index is sampled according to $\Prob{a_t^i=j} \propto w_{t-1}^j$, and $\nls_t^i$ as $\nls_t^i \sim p_\theta(\nls_t|\nls_{t-1}^{a_t^i})$. The trajectories are then augmented as $\nls_{1:t}^i = \{\nls_{1:t-1}^{a_t^i},\nls_t^i \}$.

This is repeated for $i = 1, \dots, N-1$, whereas $\nls_t^N$ is set as
$\nls_t^N = \nls'_t$. To `find' the history for $\nls_t^N$, the ancestor index $a_t^N$ is drawn with probability
\begin{align}
\Prob{a_t^N = i} \propto p_\theta(\nls_{1:t-1}^i | \nls'_{t:T}, y_{1:T}).\label{eq:as}
\end{align}
The probability density in~\eqref{eq:as} is proportional to
\begin{align}
p_\theta( y_{t:T},\nls'_{t:T}| \nls_{1:t-1}^i, y_{1:t-1}) p_\theta( \nls_{1:t-1}^i | y_{1:t-1}),
\end{align}
where the last factor is the importance weight $w_{t-1}^i$.

By sampling $\nls_{1:T}[k+1] = \nls_{1:T}^{J}$ from the rendered set of trajectories $\{ \nls_{1:T}^{i}, w_T^{i}\}_{i=1}^N$ with $\Prob{J = j} = w_T^j$, a Markov kernel $\K_\theta$ mapping $\nls_{1:T}[k] = \nls'_{1:T}$ to $\nls_{1:T}[k+1]$ is obtained.
For this Markov kernel to be useful for statistical inference
we require that \emph{(i)} it is ergodic, and \emph{(ii)} it admits $p_\theta(\nls_{1:T}|y_{1:T})$
as its unique limiting distribution. While we do not dwell on the (rather technical) details here,
we note that these requirements are indeed fulfilled; see \cite{lindsten2014particle}.

\subsection{Rao-Blackwellization}
Rao-Blackwellization of particle filters is a fusion of the Kalman filter and the particle filter based on \eqref{eq:rb}, and it is described in, e.g., \cite{schon2005marginalized}. However, Rao-Blackwellization of a particle smoother is somewhat more involved since the process $x_t|y_{1:T}$ is Markovian, but not $s_t|y_{1:T}$ (with $z_t$ marginalized, see, e.g., \cite{whiteley2010efficient} and \cite{lindsten2013backward} for various ways to handle this).

A similar problem as for the particle smoothers arises in the ancestor sampling \eqref{eq:as} in the CPF-AS. In the case of a non-Rao-Blackwellized CPF-AS, \eqref{eq:as} reduces to $w_{t-1}^i p(x'_t|x^i_{t-1})$ \cite{lindsten2014particle}. This does not hold in the Rao-Blackwellized case.

To handle this, \eqref{eq:as} can be rewritten as
\begin{align}
&w_{t-1}^i p(y_{t:T},s'_{t:T}|s^i_{1:t-1},y_{1:t-1}). \label{eq:rbas3}
\end{align}
Using the results from Section 4.4 in \cite{lindsten2013backward} (adapted to model \eqref{eq:jmlgss}), this can be written (omitting $w_{t-1}^i$, and with the notation $\|\ls\|^2_\Omega \triangleq \ls^T\Omega\ls$, $P \triangleq \Gamma\Gamma^T $, i.e. the Cholesky factorization, $Q_t \triangleq F_tF_t^T$ and $A_t \triangleq A_{s'_{t}}$ etc.)
\begin{subequations}
\begin{align}
p(y_{t:T},s'_{t:T}|s^i_{1:t-1},y_{1:t-1}) &\propto Z_{t-1}|\Lambda_{t-1}|^{-1/2}\exp(-\frac{1}{2}\eta_{t-1}),
\end{align}
with
\begin{align}
\Lambda_{t} &= \Gamma_{f;t}^{i,T}\Omega_t\Gamma_{f;t}^i+I, \label{eq:Lambdat}\\
\eta_{t} &= \|\widehat{z}^{i}_{f;t}\|^2_{\Omega_t} - 2\lambda_t^T\widehat{z}^{i,T}_{f;t} - \| \Gamma^n_{f;t}(\lambda_t-\Omega_t\widehat{z}^n_{f;t})\|^2_{M_t^{-1}}, \label{eq:etat}
\end{align}
where
\begin{align}
\Omega_{t} &= A^T_{t+1}\left(I-\widehat{\Omega}_{t+1}F_{t+1}M_{t+1}^{-1}F^T_{t+1}\right)\widehat{\Omega}_{t+1}A_{t+1} \label{eq:omegat},\\
\widehat{\Omega}_t &= \Omega_t + C_t^TR_t^{-1}C_t,\\
M_t &= F_t^T\widehat{\Omega}F_t+I,\\
\lambda_{t} &= A^T_{t+1}\left(I-\widehat{\Omega}_{t+1}F_{t+1}M_{t+1}^{-1}F^T_{t+1}\right)m_t, \label{eq:lambdat}\\
\widehat{\lambda}_t &= \lambda_t + C_t^TR_t^{-1}(y_t-D_tu_t),\\
m_t &= (\widehat{\lambda}_{t+1}-\widehat{\Omega}_{t+1}B_{t+1}u_{t+1}). \label{eq:mt}
\end{align}
and $\Omega_T = 0$ and $\lambda_T = 0$. The Rao-Blackwellization also includes an RTS smoother for finding
$p_{\theta}(\ls_{1:T}|\nls_{1:T}, y_{1:T})$.
\end{subequations}

Summarizing the above development, the Rao-Blackwellized CPF-AS (for the jump Markov linear model \eqref{eq:jmlgss}) is presented in Algorithm~\ref{alg:rbpas}, where
\begin{align}
p_\theta(y_t|&\nls_{1:t}^i,y_{1:t-1}) = \mathcal{N}(y_t; C_{s_t^i}\widehat{\ls}^n_{p;t} + D_{s_t^i}u_t,C_{s_t^i}P_{p;t}C_{s_t^i}^T + R_{s_t^i}) \label{eq:cpf1}
\end{align}
is used.
\rev{Note that the discrete state $\nls_t$ is drawn from a discrete distribution defined by $\Pi$, whereas the linear state $\ls_t$ is handled analytically.} The algorithm implicitly defines a Markov kernel $\K_\theta$ that can be used in Algorithm~\ref{alg:rbpmcmc} for finding $p(\nls_{1:T}|y_{1:T})$, or, as we will see, be placed in an SAEM framework to estimate $\theta$ (both yielding PMCMC \cite{andrieu2010particle} constructions).

\begin{algorithm}[tbh]
\begin{algorithmic}[1]
\caption{Rao-Blackwellized CPF-AS}
\label{alg:rbpas}
\REQUIRE $\nls'_{1:T} = \nls_{1:T}[k]$
\ENSURE $\nls_{1:T}[k+1]$ (A draw from $\K_\theta(\cdot | \nls_{1:T}[k]$) \emph{and} $\{\nls^{i}_{1:T}, w_T^{i} \}_{i=1}^N$
\STATE Draw $\nls^i_1 \sim p_{1}(\nls_1|y_1)$ for $i = 1, \dots, N-1$.
\STATE Compute $\{\Omega_t, \lambda_t\}_{t=1}^T$ for $\nls'_{1:T}$ according to \eqref{eq:omegat} - \eqref{eq:mt}.
\STATE Set $(\nls_1^N, \dots,  \nls_T^N) = (\nls_1', \dots, \nls_T')$.
\STATE Compute $\widehat{\ls}_{f,1}^i$ and $P_{f,1}^i$ 
 $i = 1, \dots, N$.
\STATE Set $w_1^i \propto p_\theta(y_1|\nls_1^i)$ \eqref {eq:cpf1} for $i = 1, \dots, N$ 
s.t. $\sum_i w_1^i = 1$
\FOR{$t=2$ to $T$}
\STATE Draw $a_t^i$ with $\Prob{a^i_t = j} = w_{t-1}^j$ for $i = 1, \dots, N-1$.
\STATE Draw $\nls_t^i$ with $\Prob{\nls_t^i = n} = \pi_{\nls_{t-1}^{i},n}$ for $i = 1, \dots, N-1$.
\STATE Compute $\{\Lambda_{t-1}^i,\eta_t^i\}$ according to \eqref{eq:Lambdat}-\eqref{eq:etat}.
\STATE Draw $a_t^N$ with $\Prob{a_t^N = i} \propto w^i_{t-1}\pi_{s_{t-1}^i,s_t^N}|\Lambda_{t-1}^i|^{-1/2}\exp(-\frac{1}{2}\eta_{t-1}^i)$.
\STATE Set $\nls_{1:t}^i = \{\nls_{1:t-1}^{a_t^i},\nls_t^i\}$ for $i = 1, \dots, N$.
\STATE Set $\widehat{\ls}_{f,1:t-1}^i = \widehat{\ls}_{f,1:t-1}^{a_t^i}$, $P^i_{f,1:t-1} = P_{f,1:t-1}^{a_t^i}$,
$\widehat{\ls}_{p,1:t-1}^i = \widehat{\ls}_{p,1:t-1}^{a_t^i}$ and $P^i_{p,1:t-1} = P_{p,1:t-1}^{a_t^i}$ for $i = 1, \dots, N$.
\STATE Compute $\widehat{\ls}_{p;t}^i$, $P_{p;t}^i$, $\widehat{\ls}_{f;t}^i$ and $P_{f;t}^i$ for $i = 1, \dots, N$. 
\STATE Set $w_t^i \propto p_\theta(y_t|\nls_t^i,y_{1:t-1})$ for $i = 1, \dots, N$  s.t. $\sum_i w_t^i = 1$.
\ENDFOR
\FOR{$t = T$ to $1$}
\STATE Compute $\widehat{\ls}_{s;t}^i$, $P_{s;t}^i$ for $i = 1, \dots, N$ 
\ENDFOR
\STATE Set $\nls_{1:T}[k+1] = \nls_{1:T}^{J}$ with $\Prob{J = j} = w_T^j$
\end{algorithmic}
\end{algorithm}

\section{Identification of jump Markov linear models}\label{sec:identification}

In the previous section, an ergodic Markov kernel $\K_\theta$ leaving $p_\theta(\nls_{1:T}|y_{1:T})$
invariant was found as a Rao-Blackwellized CPF-AS summarized in Algorithm~\ref{alg:rbpas}. 
This will be used together with SAEM, as it allows us to make one parameter update at each step of the Markov chain smoother in Algorithm~\ref{alg:rbpmcmc}, as presented as PSAEM in \cite{lindsten2013efficient}. (However, following \cite{lindsten2013efficient}, we make use of all the particles generated by CPF-AS, and not only $s_{1:T}[k+1]$, to compute the intermediate quantity in the SAEM.)

This leads to the approximation (cf.\ \eqref{eq:Qsaem})
\begin{align}
  \nonumber
  &\saemQ_k(\theta) = (1-\gamma_k)\saemQ_{k-1}(\theta) + \\
  \label{eq:Qsaem2}
  &\gamma_k \sum_{i=1}^N w_{T}^i \Expb{\theta_k}{ \log p_{\theta}(y_{1:T}, z_{1:T}, s_{1:T}^{i}) | s_{1:T}^{i}, y_{1:T}},
\end{align}
where the expectation is w.r.t.\ $z_{1:T}$.
Putting this together, we obtain a Rao-Blackwellized PSAEM (RB-PSAEM)
algorithm presented in Algorithm~\ref{alg:rbpsaem}. Note that
this algorithm is similar to the MCMC-based smoother in Algorithm~\ref{alg:rbpmcmc},
but with the difference that the model parameters are updated at each iteration,
effectively enabling simultaneous smoothing and identification.

\begin{algorithm}
\caption{Rao-Blackwellized PSAEM}
\label{alg:rbpsaem}
\begin{algorithmic}[1]
\STATE Initialize $\widehat{\theta}_0$ and $\nls_{1:T}[0]$, and ${\saemQ}_0(\theta) \equiv 0.$
\FOR{$k \geq 1$}
\STATE Run Algorithm~\ref{alg:rbpas} to obtain $\{\nls^{i}_{1:T},w_T^{i} \}_{i=1}^N$\\and $\nls_{1:T}[k]$.
\STATE Compute ${\saemQ}_k(\theta)$ according to \eqref{eq:Qsaem2}.
\STATE Compute $\widehat{\theta}_k = \arg \max_{\theta \in \Theta}{\saemQ}_k(\theta)$ \label{algstep:mq}
\ENDFOR
\end{algorithmic}
\end{algorithm}

(For notational convenience, the iteration number $k$ is suppressed in the variables related to $\{\nls^{i}_{1:T}, w_T^{i} \}_{i=1}^N$.)

With a strong theoretical foundation in PMCMC and Markovian stochastic approximation, the RB-PSAEM algorithm
presented here enjoys very favourable convergence properties. In particular, under certain smoothness and ergodicity conditions, the sequence
of iterates $\{\theta_k\}_{k\geq 1}$ will converge to a maximizer of $p_\theta(y_{1:T})$ as $k\rightarrow\infty$,
regardless of the number of particles $N\geq2$ used in the internal CPF-AS procedure (see \cite[Proposition~1]{lindsten2013efficient} together with~\cite{kuhn2004coupling} for details).
Furthermore, empirically it has been found that a small number of particles can work well in practice as well.
For instance, in the numerical examples considered in Section~\ref{sec:numerical}, we run Algorithm~\ref{alg:rbpsaem}
with $N = 3$ with accurate identification results.

For the model structure \eqref{eq:jmlgss}, there exists infinitely many solutions to the problem \eqref{eq:ML}; all relevant involved matrices can be transformed by a linear transformation matrix and the modes can be re-ordered, but the input-output behaviour will remain invariant. The model is therefore over-parametrized, or lacks identifiability, in the general problem setting. However, it is shown in \cite{pintelon1996minimum} that the Cram\'{e}r-Rao Lower Bound is not affected by the over-parametrization. That is, the estimate quality, in terms of variance, is unaffected by the over-parametrization.

\subsection{Maximizing the intermediate quantity}

When making use of RB-PSAEM from Algorithm~\ref{alg:rbpsaem}, one major question arises from Step~\ref{algstep:mq}, namely the maximization of the intermediate quantity $\saemQ_k(\theta)$. For the jump Markov linear model, the expectation in~\eqref{eq:Qsaem2} can be expressed using sufficient statistics, as will be shown later, as an inner product
\begin{align}
	&\sum_{i=1}^N w_{T}^i \Expb{\theta_k}{ \log p_{\theta}(y_{1:T}, z_{1:T}, s_{1:T}^{i}) | s_{1:T}^{i}, y_{1:T}} = 
	\langle S^k,\eta(\theta) \rangle,  \label{eq:inp}
\end{align}
for a sufficient statistics $S$ and corresponding natural parameter $\eta(\theta)$. Hence $\saemQ_k$ can be written as
\begin{align}
\saemQ_k(\theta) = (1-\gamma_k)\saemQ_{k-1}(\theta) + \gamma_k\langle S^k,\eta(\theta) \rangle = \langle \saemS^k,\eta(\theta) \rangle \label{eq:max1}
\end{align}
if the transformation
\begin{align}
\saemS^k = (1-\gamma_k)\mathbb{S}^{k-1} + \gamma_kS^k \label{eq:sasufstat}
\end{align}
is used. In detail, 
\begin{subequations}
\begin{align}
&\sum_{i=1}^N w_{T}^i \Expb{\theta_k}{ \log p_{\theta}(y_{1:T}, z_{1:T}, s_{1:T}^{i}) | s_{1:T}^{i}, y_{1:T}} = \nonumber\\
&\sum_{n=1}^K\sum_{m=1}^KS_{n,m}^{(1)}\log \pi_{n,m} 
- \sum_{n=1}^K \frac{1}{2}\left(S^{(2)}_{n}\log(|Q_n||R_n|) + \Tr(H_n^\theta S^{(3)}_{n})\right) \label{eq:max4}
\end{align}
neglecting constant terms in the last expression. This can be verified to be an inner product (as indicated in \eqref{eq:inp}) in $S = \{S^{(1)},S^{(2)},S^{(3)}\}$. Here the sufficient statistics
\begin{align}
S^{(1)}_{n,m} &= \sum_{i = 1}^N w^i_T \sum_{t=1}^T \Ind{\nls^{i}_t = m, \nls_{t-1}^i = n},\label{eq:sufstat1}\\
S^{(2)}_{n} &= \sum_{i = 1}^N w^i_T \sum_{t=1}^T   \Ind{\nls^{i}_t = n} ,\label{eq:sufstat2}\\
S^{(3)}_{n} &= \sum_{i = 1}^N w^i_T \sum_{t=1}^T   \Ind{\nls_t^{i} = n}(\widehat{\xi}^{i}_t\widehat{\xi}_t^{i,T}+M_{t|T}^i),\label{eq:sufstat3}
\end{align}
with
\begin{align}
\widehat{\xi}^i_t &= \left( \widehat{\ls}_{s;t}^{i,T} ~ \left[ \widehat{\ls}_{s;t-1}^{i,T} ~ u_{t-1}^{T} \right] ~ y_t^{T} ~ \left[ \widehat{\ls}_{s;t}^{i,T} ~u_t^T \right] \right)^T, \label{eq:xihat}
\end{align}
and
\begin{align}
H_n^\theta &= \begin{pmatrix}
\begin{bmatrix} I & A_n^T & B_n^T \end{bmatrix} Q_n^{-1} \begin{bmatrix} I \\ A_n \\ B_n \end{bmatrix} & 0 \\
0 & \begin{bmatrix} I & C_n^T & D_n^T \end{bmatrix} R_n^{-1} \begin{bmatrix} I \\ C_n \\ D_n \end{bmatrix}
\end{pmatrix}
\end{align}
have been used. Further notation introduced is $\Ind{\cdot}$ as the indicator function, and 
\begin{align}
  &M_{t|T}^i = \begin{pmatrix}
 P^i_{s;t} & P^i_{s;t,t-1} & 0 & 0 & P^i_{s;t} & 0\\
 P^i_{s;t,t-1} & P^i_{s;t-1} & 0 & 0 & P^i_{s;t,t-1} & 0 \\
 0 & 0 & 0 & 0 & 0 & 0\\
 0 & 0 & 0 & 0 & 0 & 0\\
 P^i_{s;t} & P^i_{s;t,t-1} & 0 & 0 & P^i_{s;t-1} & 0 \\
 0 & 0 & 0 & 0 & 0 & 0 \end{pmatrix}.
\end{align}
\end{subequations}

For computing this, the RTS-smoother in step 17 in Algorithm~\ref{alg:rbpas} has to be extended by calculation of $P_{s;t+1,t} \triangleq \Cov{\widehat{\ls}_{s,t+1}\widehat{\ls}_{s;t}^T}$, which can be done as follows \cite[Property~P6.2]{sumway2006time}
\begin{align}
  \label{eq:ksP2}
  P_{s;t,t-1} = P_{f;t}J_{t-1}^T + J_t(P_{s;t+1,t}-A_{t+1}P_{f;t})J_{t-1}^T, 
\end{align}
initialized with $P_{T,T-1|T} = (I - K_TC_T)A_TP_{f;t-1}$.

For notational convenience, we will partition $S^{(3)}_n$ as
\begin{align}
S^{(3)}_n = \begin{pmatrix}
 {\Phi}_n & {\Psi}_n &  &  \\
 {\Psi}_n^T & {\Sigma}_n &  &  \\
  &  & {\Omega}_n & {\Lambda}_n \\
  &  & {\Lambda}_n^T & {\Xi}_n
 \end{pmatrix}.\label{eq:part}
\end{align}

\begin{lemma}\label{lemma:1}
Assume for all modes $n = 1, \dots, K$, that all states $\ls$ are controllable and observable and $\sum_t \Ind{\nls_t = n} u_t^Tu_t > 0$.
The parameters $\theta$ maximizing $\saemQ_k(\theta)$ for the jump Markov linear model \eqref{eq:jmlgss} are then given by
\begin{subequations}\label{eq:estAll}
\begin{align}
{\pi}^j_{n,m} &= \frac{{\saemS}^{(1),k}_{n,m}}{\sum_l {\saemS}^{(1),k}_{n,l}},\\
\begin{bmatrix} A_n & B_n \end{bmatrix} &= {\Psi}_n{\Sigma}^{-1}_n, \label{eq:estAB}\\
\begin{bmatrix} C_n & D_n \end{bmatrix} &= {\Lambda}_n{\Xi}^{-1}_n,\\
\begin{bmatrix} Q_n \end{bmatrix} &= ({\saemS}^{(2),k}_{n})^{-1}\left({\Phi}_n - {\Psi}_n{\Sigma}_n^{-1}{\Psi}_n^T\right), \label{eq:estQ}\\
\begin{bmatrix} R_n \end{bmatrix} &= ({\saemS}^{(2),k}_{n})^{-1}\left({\Omega}_n - {\Lambda}_n{\Xi}_n^{-1}{\Lambda}_n^T\right), \label{eq:estR}
\end{align}
for $n,m = 1, \dots, K$.
\end{subequations}
\end{lemma}

${\Phi}_n, {\Psi}_n, \dots$ are the partitions of $\saemS^{(3),k}_n$ indicated in \eqref{eq:part}, and $\saemS^{(i)}$ are the `SA-updates' \eqref{eq:sasufstat} of the sufficient statistics \eqref{eq:sufstat1}-\eqref{eq:sufstat3}.

\emph{Remark:} If $B \equiv 0$, the first square bracket in \eqref{eq:xihat} can be replaced by $\begin{bmatrix} \widehat{\ls}_{s;t-1}^{i,T} \end{bmatrix}$, and \eqref{eq:estAB} becomes $\begin{bmatrix} A_n \end{bmatrix} = {\Psi}_n{\Sigma}^{-1}_n$. The case with $D \equiv 0$ is fully analogous.

\begin{proof}
With arguments directly from~\cite[Lemma 3.3]{gibson2005robust}, the maximization of the last part of \eqref{eq:max4} for a given $\nls_t = n$ 
(for any sufficient statistics $Z$ in the inner product, and in particular $Z = \saemS^k$), is found to be \eqref{eq:estAB}-\eqref{eq:estR}.

Using Lagrange multipliers and that $\sum_i \pi_{n,m} = 1$, the maximum w.r.t. $\Pi$ of the first part of \eqref{eq:max4} is obtained as
\begin{align}
\pi_{n,m} = \frac{\saemS_{n,m}^{(1),k}}{\sum_l \saemS_{n,l}^{(1),k}}.
\end{align}
\end{proof}

\subsection{Computational complexity}
Regarding the computational complexity of Algorithm~\ref{alg:rbpsaem}, the most important result is
that it is linear in the number of measurements $T$. It is also linear in the number of particles
$N$.

\section{Numerical examples}\label{sec:numerical}
Some numerical examples are given to illustrate the properties of the Rao-Blackwellized PSAEM
algorithm. The Matlab code for the examples is available via the homepage of the first author.

\subsection{Example~1 - Comparison to related methods}
\begin{figure}
        \centering
		\includegraphics[width=0.6\textwidth]{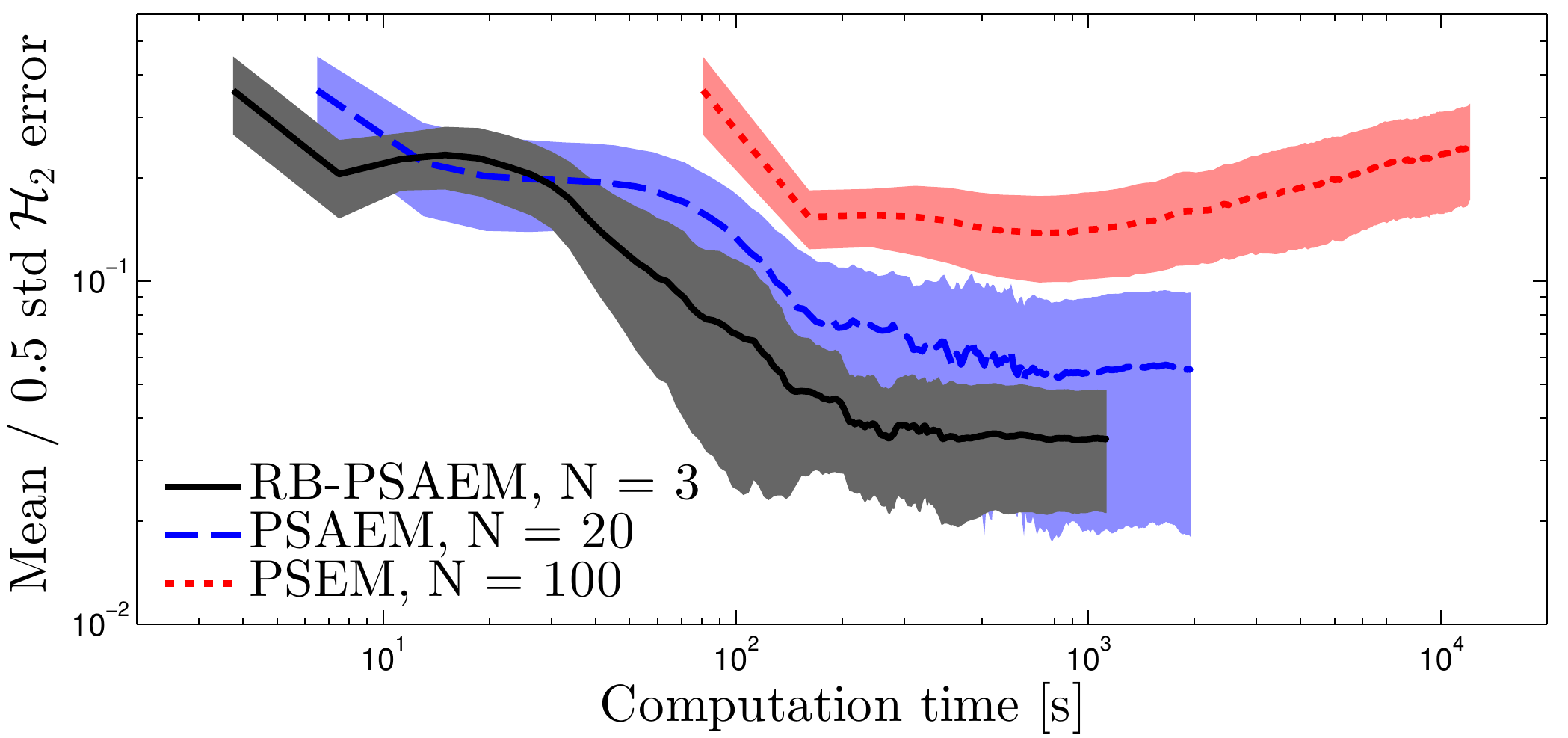}
		\caption{Numerical example 1. Mean (lines) and 0.5 standard deviation (fields) $\mathcal{H}_2$ error for 7 runs of our RB-PSAEM using $N=3$ particles (black)
                  PSAEM \cite{lindsten2013efficient} using $N=20$ particles (blue) and PSEM
                  \cite{SchonWN:2011} using $N=100$ particles and $M=20$ backward trajectories
                  (red).}
		\label{fig:num_ex1}
\end{figure}

The first example concerns identification using simulated data ($T = 3\thinspace000$) for a
one-dimensional ($\nx = 1$) jump Markov linear model with 2 modes ($K = 2$) (with parameters
randomly generated according to $A_n \sim U_{[-1,1]},~B_n \sim U_{[-5,5]},~C_n \sim U_{[-5,5]}, ~D_n
\equiv 0, Q_n \sim U_{[0.01,0.1]}, R_n \sim U_{[0.01,0.1]}$) with low-pass filtered white noise as
$u_t$. The following methods are compared:
\begin{enumerate}
\item RB-PSAEM from Algorithm~\ref{alg:rbpsaem}, with (only) $N = 3$ particles,
\item PSAEM as presented in \cite{lindsten2013efficient} with $N = 20$,
\item PSEM \cite{SchonWN:2011} with $N = 100$ forward particles and $M = 20$ backward simulated trajectories.
\end{enumerate}
The initial parameters $\widehat{\theta}_0$ are each randomly picked from
$[0.5\theta^\star,1.5\theta^\star]$, where $\theta^\star$ is the true parameter value. The results are
illustrated in Figure~\ref{fig:num_ex1}, which shows the mean (over all modes and $7$ runs)
$\mathcal{H}_2$ error for the transfer function from the input $u$ to the output $y$.

From Figure~\ref{fig:num_ex1} (note the log-log scale used in the plot) it is clear that our new Rao-Blackwellized PSAEM algorithm has a significantly better performance, both in terms of mean and in variance between different runs, compared to the previous algorithms.

\subsection{Example~2 - Identification of multidimensional systems}
\begin{figure}[tb]
	\centering
	\begin{subfigure}[b]{0.47\textwidth}
		\centering
		\includegraphics[width=\textwidth]{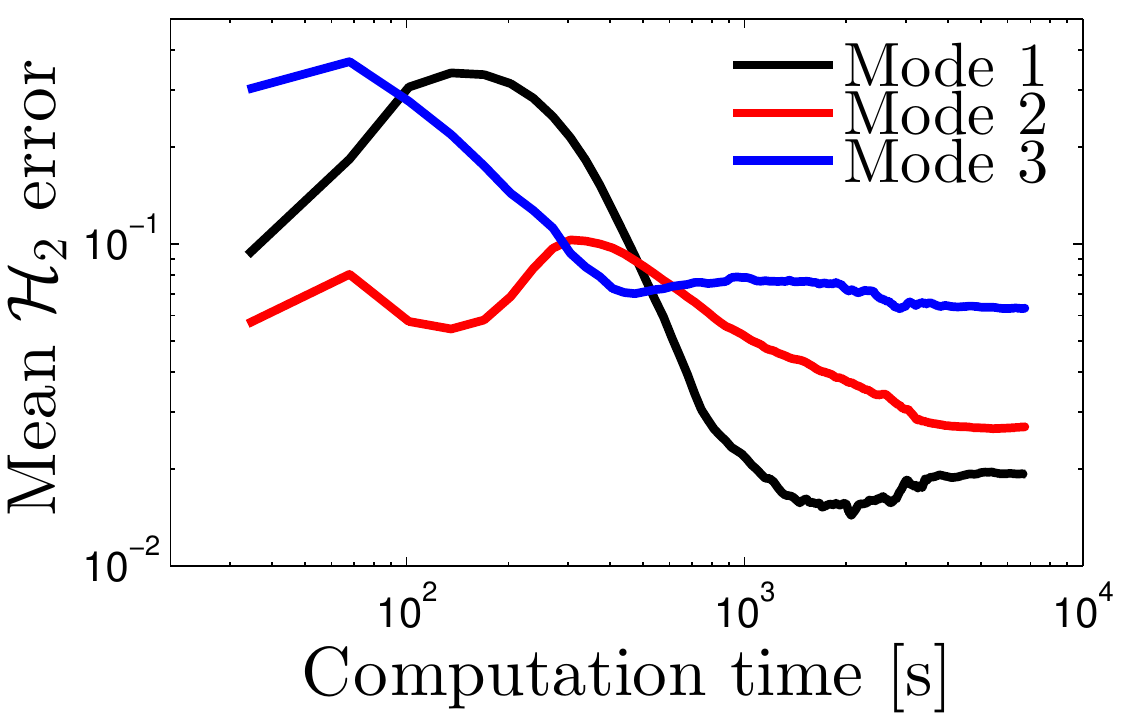}
		\caption{Mean $\mathcal{H}_2$ error for each mode.}
		\vspace{0.35cm}
		\label{fig:num_ex2}
	\end{subfigure}
	~
	\begin{subfigure}[b]{0.47\textwidth}
		\centering
		\includegraphics[width=\textwidth]{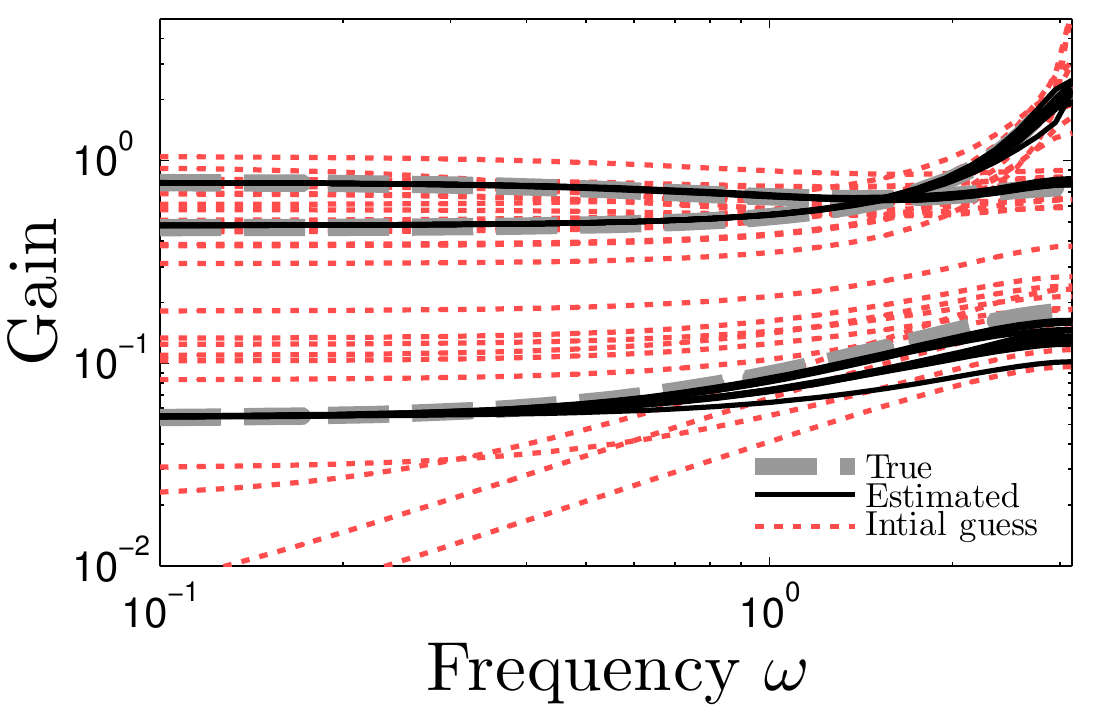}
		\caption{Bode plots of the estimates (black), true (dashed grey) and the initializations
                  (dotted red).}
		\label{fig:num_ex2_3}
	\end{subfigure}
	\caption{Plots from Numerical example 2.}
\end{figure}

Let us now consider a two-dimensional system ($\nx = 2$) with $K = 3$ modes. The eigenvalues for
$A_n$ are randomly picked from $[-1,1]$. The other parameters are randomly picked as $B_n \sim
U_{[-5,5]},~C_n \sim U_{[-5,5]}, ~D_n \equiv 0, Q_n \sim I_2 \cdot U_{[0.01,0.1]}, R_n \sim
U_{[0.01,0.1]},$ and the system is simulated for $T = 8\thinspace 000$ time steps with input $u_t$
being a low-pass filtered white noise. The initialization of the Rao-Blackwellized PSAEM algorithm
is randomly picked from $[0.6\theta^\star,1.4\theta^\star]$ for each parameter. The number of
particles used in the particle filter is $N = 3$. Figure~\ref{fig:num_ex2} shows the mean (over $10$
runs) $\mathcal{H}_2$ error for each mode, similar to
Figure~\ref{fig:num_ex1}. Figure~\ref{fig:num_ex2_3} shows the estimated Bode plots after $300$
iterations. As is seen from Figure~\ref{fig:num_ex2_3}, the RB-PSAEM algorithm has the ability to catch the dynamics of the multidimensional system fairly well.

\section{CONCLUSION AND FUTURE WORK}
We have derived a maximum likelihood estimator for identification of jump Markov linear models. More
specifically an expectation maximization type of solution was derived. The nonlinear state smoothing
problem inherent in the expectation step was solved by constructing an ergodic Markov kernel leaving
the joint state smoothing distribution invariant. Key to this development was the introduction of a
Rao-Blackwellized conditional particle filter with ancestor sampling. The maximization step could be
solved in closed form. The experimental results indicate that we obtain significantly better
performance both in terms of accuracy and computational time when compared to previous state of the
art particle filtering based methods. The ideas underlying the smoother derived in this work have
great potential also outside the class of jump Markov linear models and this is something worth more
investigation. Indeed, it is quite possible that it can turn out to be a serious competitor also in
finding the joint smoothing distribution for general nonlinear state space models.

\bibliographystyle{plain}
\bibliography{refs}

\end{document}